\documentclass{aa}
\usepackage{txfonts}
\usepackage{graphicx}
\usepackage{natbib}
\bibpunct{(}{)}{;}{a}{}{,}

\begin{document}
\title{Hard X-ray Spectra and Positions of Solar Flares observed by RHESSI:
photospheric albedo, directivity and electron spectra}

   \author{Jana Ka\v{s}parov\'{a} \inst{1}
          \and
          Eduard P. Kontar \inst{2}
	\and 
	 John C. Brown \inst{2}  
          }

   \offprints{Jana Ka\v{s}parov\'{a}}

   \institute{Astronomical Institute, Academy of Sciences of the Czech Republic,
Fri\v{c}ova 298, 251 65 Ond\v{r}ejov, Czech Republic \email{kasparov@asu.cas.cz}
          \and
   Department of Physics and Astronomy, University of Glasgow,
G12 8QQ, UK \\ \email{eduard,john@astro.gla.ac.uk}
             }

   \date{Received *, 2006; accepted *, 2006}

\abstract
 {}
 {We investigate the signature of the photospheric albedo contribution in solar flare
  hard X-ray spectra, the effect of low energy cutoffs in electron spectra, and the 
  directivity of hard X-ray emission.}
 {Using Ramaty High Energy Solar Spectroscopic Imager (RHESSI) flare data
  we perform a statistical analysis of spatially integrated spectra and positions of solar flares.}
 {We demonstrate clear centre-to-limb variation of photon spectral indices
   in the 15~-~20 keV energy range and a weaker dependency in the 20~-~50 keV range which
   is consistent with photospheric albedo as the cause. The results also suggest that low-energy cutoffs
   sometimes inferred in mean electron spectra are an artefact of albedo.
   We also derive the anisotropy (ratio of downward/observer directed photons) of hard X-ray emission
    in the 15~-~20~keV range for various heliocentric angles.
   }	
 {}

\titlerunning{Photospheric albedo and directivity}
   \maketitle

\section{Introduction}

The successful operation of the Ramaty High Energy Solar Spectroscopic
Imager {\it RHESSI} \citep{li02} since February 2002
has provided us with a substantial database of solar hard X-ray flares that
can be used for detailed spectroscopic analysis. The high accuracy
of the spectroscopic data, in combination with the positions of solar
flares, allows us to determine the centre-to-limb spectral variations in
various energy ranges.

Variation of hard X-ray flare occurrence and spectral index with solar position of
the sources have been of great interest for several decades.
As suggested e.g. by %
\citet[][]{ha71,br72a,pe73,le83}, centre-to-limb variations could provide essential
information on directivity of X-ray emission and hence accelerated electrons.
Most earlier analyses concentrated on centre-to-limb variation of flare occurrence.
\citet{oh69} and \citet{pi69} reported significant centre-to-limb variation at energies above 10~keV
although they did not correct the hard X-ray flare occurrence for H$\alpha$ flare 
longitude distribution. Subsequent works \citep{dr71,ca73,ph73,da74,ka74} 
covering various energy ranges from 1~-~10~keV claimed no centre-to-limb 
variation of hard X-ray occurrence. 
\citet{da77} concluded that anisotropy in the 10~-~100~keV range was
so low that it ruled out the downward-beamed thick target model 
due to lack of limb brightening predicted e.g. by \citep{br72a}.
Theoretical models \citep{pe73,la77,ba78} predicted
that at hard X-ray energies, above 15~keV, centre-to-limb variations of spectral properties
should be sensitive to electron directivity.
Although such studies were not very common, 
\citet{da74} and \citet{ro75} found, contrary to above mentioned works based on hard X-ray occurrence, significant 
centre-to-limb spectral variation in
the 10~-~100~keV range and attributed it to anisotropy of X-ray emission. Then,
\citet{ve87} demonstrated that Solar Maximum Mission (SMM) GRS data
in the 25~-~200~keV range and above 300~keV,
show significant spectral hardening of events near the limb.
Similar results were reported
by \citet{bo85} from Venera 13 data in the 50~-~100~ keV range and by \citet{mc91}
in the 0.3~-~1~MeV range.

It has been known for a while that X-ray photons can
be effectively backscattered by photosphere atoms and electrons 
\citep{to72,ba78}. Thus, at energies not dominated by absorption
the backscattered albedo flux must be
seen virtually in every solar flare spectrum, the degree of the
albedo contribution depending on the directivity of the primary X-ray flux
\citep{ko06}. The solar flare photons backscattered by the solar
photosphere {\bf can} contribute significantly ({\bf the reflected flux is}
50-90 \% {\bf of the primary} in the 30~-~50~keV
range for isotropic sources) to the total observed
photon spectrum. For the simple case of a power-law-like primary solar
flare spectrum (without albedo), the photons reflected by the photosphere produce 
a broad `hump' component.
Photospheric albedo makes the
observed spectrum flatter below $\sim 35$ keV and slightly steeper above,
in comparison with the primary spectrum.
The amount of backscattered photons is {\bf given by the scattering cross-section and is} 
roughly dependent on the projected area of the albedo patch, which makes the photospheric albedo
stronger for solar centre events. Photospheric albedo has
been treated using Monte Carlo simulations to calculate the observed
spectrum for an assumed primary power-law spectrum \citep{ba78}.
\citet{ko06} developed a new approach to photospheric
albedo based on the Green's functions of \citet{ma95} that produces
the correction for any solar flare X-ray spectrum and
can be used both for forward and inverse spectrum calculations.
The influence of albedo was not taken into account for spectral
centre-to-limb variation in earlier analyses and it was wrongly
dismissed \citep{da74} as being unimportant in the 20~-~50~keV
range. However, \citet{ba78} and \citet{ve87} pointed out that
the Datlowe's results are consistent with an albedo contribution to
the observed photon spectrum.
Using the \citet{ba78} results, the importance of albedo in electron spectrum inference was
first included by \citet{jo92} and analysed by \citet{al02}.

\begin{figure}
\begin{center}
\includegraphics[width=80mm]{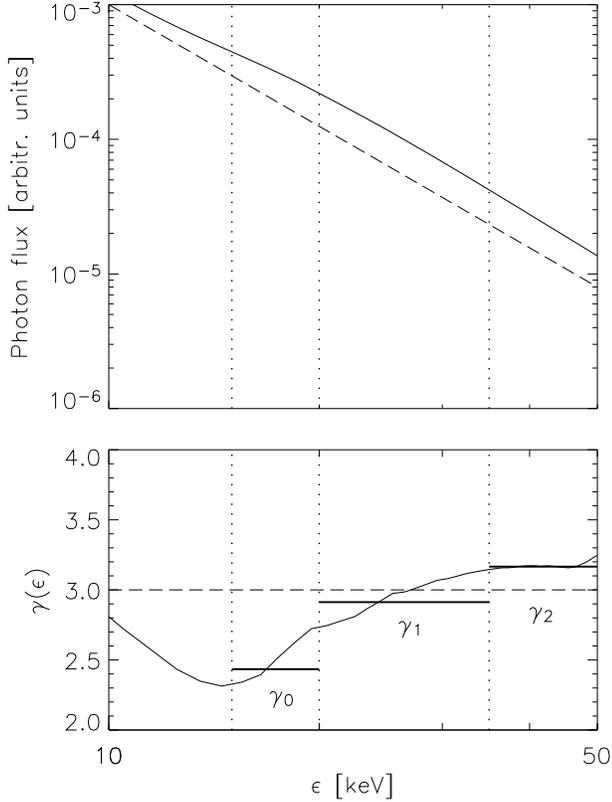}
\end{center}
\caption{Photon spectrum $I(\epsilon)$ (upper panel) and spectrum
index $\gamma (\epsilon)$ (lower panel) for a modelled primary
photon spectrum $I_\mathrm{p}(\epsilon)\sim \epsilon ^{-3}$
for an {\bf isotropic} source located at $\mu=0.8$. Primary 
$I_\mathrm{p}(\epsilon)$ and $\gamma_\mathrm{p} (\epsilon)$ are plotted with a dashed line and the  observed
$I_\mathrm{o}(\epsilon)$ and $\gamma_\mathrm{o} (\epsilon)$ with a solid line.
Horizontal lines indicate spectral indices $\gamma_0$, $\gamma_1$, and $\gamma_2$ 
obtained by single power-law fits in the 15~-~20, 20~-~35, and 35~-~50~keV energy ranges, 
respectively.}
\label{model}
\end{figure}

On the other hand, it has been long believed that the mean electron flux distribution
$\bar{F}(E)$ defined by \citep{br71,br03} can exhibit low-energy cutoff -- e.g.
\citet{ni90} showed that photon spectra of the impulsive
component flatten toward low energies and suggested
that a low-energy cutoff of the electron spectrum around 50 keV
could be responsible. \citet{fa97}, using Yohkoh data, observed a flare
with a flat spectrum below 33 keV (photon spectral index as low as 1.98).
More recently, \citet{ka05b} reported a clear
low-energy cutoff inferred from the observed photon spectra of
the August 20, 2002 flare, neglecting albedo. %
Using {\bf a regularised} inversion technique, \citet{ko06b}
report three flares with a similar feature -- an absence of electrons 
around 20~-~30 keV between possibly thermal and non-thermal
component. 

However, applying the model-independent albedo correction method
of \citet{ko06} to the August 20, 2002 flare, \citet{ka05b} have
clearly demonstrated that such a high value of low-energy cutoff is
removed by albedo correction for an isotropic source.
Equally, the albedo corrected photon spectra do {\it not}
require a gap in the inverted electron spectrum
\citep{ka05b,ko06b}. Using the present survey, we can potentially find 
many flares with such low spectral indices that they might have clear
low-energy cutoff appearing as a gap between thermal and non-thermal
distributions. 

The spectrum-independent photospheric albedo correction method is
introduced in Section 2. Data selection and analysis are discussed
in Section 3. In Section 4, our survey shows that the
centre-to-limb variations  of spectral indices are consistent with
photospheric albedo. Moreover, we find limits on the X-ray spectral directivity for
various heliospheric angles. Section 5 confirms the prediction that 
the low-energy cutoff in $\bar{F}(E)$ is removed
when the spectra are albedo corrected for all events with low
spectral index found in the survey. Lastly, Section 6 reviews the
results obtained and their consequences for the angular distribution
of X-ray emitting electrons.

\section{Photospheric albedo and spectral index}
\label{sect_albedo}
The albedo contribution produces substantial change to the observed spectrum in
the range between 10 and 100 keV. It becomes negligible for energies below 10 keV due 
to the photoelectric absorption by atoms while photons above 100 keV 
penetrate deep into high density layers of the solar atmosphere 
and do not escape back. Therefore, the reflectivity of the photosphere has a
broad maximum in the 30~-~50 keV range. The total spectrum
in the observer direction $\mu$
($\mu=\cos\theta$, where $\theta$ is the heliocentric angle)
can be expressed as
\begin{displaymath}
    I_\mathrm{o}(\epsilon,\mu) = I_\mathrm{p} (\epsilon,\mu) + I_\mathrm{A} (\epsilon,\mu)
\end{displaymath}
\begin{equation}\label{flux0}
    I_\mathrm{A} (\epsilon,\mu)={\bf G} (\epsilon,\epsilon',\mu) I_\mathrm{p} (\epsilon',\mu) \alpha (\mu)\ ,
\end{equation}
where $I_\mathrm{p}(\epsilon,\mu)$ is the primary photon spectrum of the flare,
$I_\mathrm{A} (\epsilon,\mu)$ is the spectrum reflected into the observer direction, 
${\bf G}$ is the Green's matrix \citep{ko06} accounting for
Compton backscattering and photoelectric absorption, and
$\alpha (\mu)$ is a parameter of emission anisotropy, 
roughly
the ratio of the flux toward the Sun $I_\mathrm{p} (\epsilon,\mu<0)=\alpha (\mu)I_\mathrm{p} (\epsilon,\mu)$ and
the flux toward the observer $I_\mathrm{p} (\epsilon,\mu)$. 
We simplify the problem here by considering only $\alpha(\mu)$ 
independent of $\epsilon$ \citep[cf][]{al06}.

The photon spectrum, $I_\mathrm{A} (\epsilon,\mu)$, backscattered toward 
the observer strongly depends on three parameters:
the spectral index of the primary spectrum, the location of the
flare on the solar disc, and the directivity of X-ray emission
(anisotropy parameter $\alpha(\mu)$). The contribution of the
albedo to the observed spectrum is greater for flatter primary
spectra (low $\gamma$) and for solar disc flares (large
$\mu$). Further, the more  X-rays are beamed downwards $\alpha(\mu)> 1$, the larger
is the contribution of the albedo to the observed spectrum.

The energy dependent spectral index
\begin{equation}\label{gamma}
    \gamma (\epsilon) \equiv -
    \frac{\epsilon}{I(\epsilon)}\frac{\mathrm{d}I(\epsilon)}{\mathrm{d}\epsilon} = -
    \frac{\mathrm{d} \ln (I(\epsilon))}{\mathrm{d} \ln(\epsilon)}
\end{equation}
for the primary spectrum $\gamma_\mathrm{p}(\epsilon)$ substantially differs
from that ($\gamma_\mathrm{o}(\epsilon)$) of the total $I_\mathrm{o}(\epsilon,\mu)$ toward the observer 
(including albedo) -- see Figure 1 for the case of $\gamma_\mathrm{p}=3$. In the
range 10~-~25 keV the photon spectrum becomes flatter, $\gamma_\mathrm{o} <
\gamma_\mathrm{p}$, while the observed spectral index is higher at 
energies above 35 keV, $\gamma_\mathrm{o} > \gamma_\mathrm{p}$.

Using a single power-law fit in various energy ranges (Figure~\ref{model}), we can
estimate the local spectral index {\bf assuming it is constant in each energy range}. 
(Note that generally for a photon
spectrum $I(\epsilon) \sim \epsilon ^{-c (\epsilon)}$,
$c(\epsilon) = \gamma (\epsilon)$ only for $c = \mbox{const}$ \citep{co03}.)

Hence, as a result of heliocentric angle dependent albedo, 
the spectral shape of photon spectra should vary as a
function of their position on the solar disc, i.e. a
centre-to-limb variation of spectral indices is expected and the
degree of the spectral index variation with heliocentric angle
could usefully provide us with the directivity of hard X-ray
sources, i.e. the anisotropy coefficient $\alpha (\mu)$.

\section{Data selection and analysis}

In our analysis we used the whole RHESSI flare list (from February 12,
2002 until April 18, 2006) to search for any evidence of
centre-to-limb variation in photon spectra.

RHESSI has 9 detectors on board with FWHM resolution
about $\sim 1$ keV. For X-ray spectroscopy analysis we use the six
front segments of the detectors ignoring detector 2, 5 and 7 due
to either low resolution or the segmentation problem \citep{sm02}.

To handle such a large set of data, we set up an automatic procedure
which selects suitable events and retrieves the flare photon
spectra. To be chosen for further analysis the events had to meet
the following major criteria:
\begin{itemize}
\item The spectra of the events should be observed above $25$~keV.
Thus, we have selected flares flagged as observed above that
energy from the RHESSI flare list and have determined the time of
peak flux in the 25~-~50 keV range. The events flagged as particle
events were ignored. This led us to $\sim 1500$ events.

\item Count rates exceeding about $2\times 10^3$ counts
per second per detector cause pulse pile-up in the detectors that
requires troublesome correction \citep{sm02}. To avoid the
pile-up issues, we have thrown away events with corrected
livetime {\bf counter} $< 90 \%$. 
We disregarded also weak events (background subtracted
count rate lower than $5$ counts per second in the 25~-~50 keV range)
and events possibly affected by X-ray absorption in Earth's atmosphere 
(peak time closer than 60~s to RHESSI eclipse time). 
This reduces our survey to $\sim 800$ events.

\item We have used CLEANed images in the range 10~-~20 and 20~-~50~keV to
check for true solar events and to find their positions. Finally, the total number
of the selected flare events drops to 703. 
\end{itemize}

\begin{figure}
\begin{center}
\includegraphics[width=80mm]{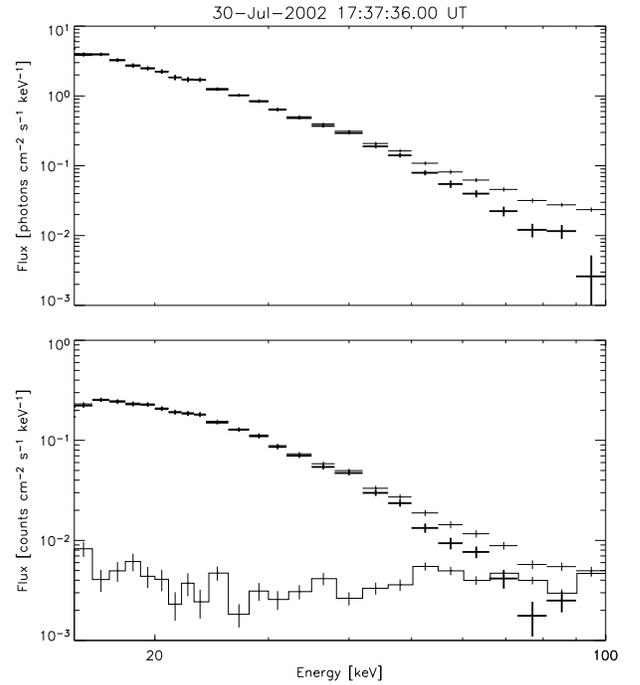}
\end{center}
\caption{Example of the flare spectrum for July 30, 2002 flare,
17:37:36--17:37:48 UT. The upper panel corresponds to the regularised photon spectrum,
the lower panel shows the count spectrum. 
Observed spectrum is plotted with thin crosses, background subtracted flare spectrum 
with thick crosses. The background is displayed as a histogram.
Horizontal sizes of crosses and width of the histogram bins correspond to widths of energy bands,
while vertical sizes represent $1\sigma$ statistical uncertainties of the flux.
}
\label{flare_example}
\end{figure}

RHESSI is an instrument with high background that makes background
subtraction an important step in the data analysis \citep{sch02}.
The code searches for two time intervals, one before
and one after the flare peak, where count rates are less than 0.01
of the peak count rate in 15~-~50 keV range and 0.1 above 50 keV.
The background time intervals are determined for several energy
bands separately and the background at the peak is estimated by a
linear fit \citep{sch02}. See the flare example in Figure~\ref{flare_example}.

The count spectra [counts sec$^{-1}$~keV$^{-1}$~cm$^{-2}$] were
accumulated in 12~s time intervals with a quasi-logarithmic energy
binning from 10 to 100~keV to improve the signal-to-noise ratio (SNR) at
higher energies (Figure~\ref{flare_example}). 
{\bf After background subtraction, }
the resulting peak count flux {\bf (one spectrum per flare)} 
and the detector response matrix (DRM) for the instrument at this time
interval were used as an input for the regularised inversion
routine \citep{sc06,ko04}. The regularised inversion of a count
flux spectrum gave us a non-parametric photon spectrum and its uncertainties.

\subsection{Photon spectra of selected events}

As discussed in Section~\ref{sect_albedo}, addition of the albedo contribution
changes the shape of the primary spectra. To describe its effects,
the regularised photon spectra have been fitted with a single power-law in
three separate energy ranges 15~-~20, 20~-~35, and 35~-~50~keV
thus obtaining corresponding spectral indices $\gamma_0$,$\gamma_1$, and
$\gamma_2$, respectively.

{\bf Note that spectra in the 15-20 keV energy range may be affected by thermal component(s) which could
be comparable to or dominate the non-thermal emission and masking the albedo contribution.
Considering a widely used isothermal approximation, value of $\gamma_0$ in such flares 
would be higher in comparison with events without strong thermal component.
}

Since the count flux decreases with energy while the background level
does not (Figure \ref{flare_example}), naturally not all 703 
events have significant
flux above background at all three energy ranges. For $\gamma_0$ and $\gamma_1$
analysis we accepted only those events with {\bf count flux  signal-to-noise ratio (SNR)}~$>3$ 
at all three energy ranges. Furthermore, the Green's matrices and 
therefore the corrected DRM are non-diagonal at
$\gamma_2$ energies, the photon flux at this energy range
depending on the signal at higher energies and making $\gamma_2$
unreliable for flares with low SNR 
above 50~keV. Thus, for $\gamma_2$ analysis additional condition  
SNR~$>3$ in the 50~-~66~keV energy range was required.

\begin{figure}
\begin{center}
\includegraphics[width=85mm]{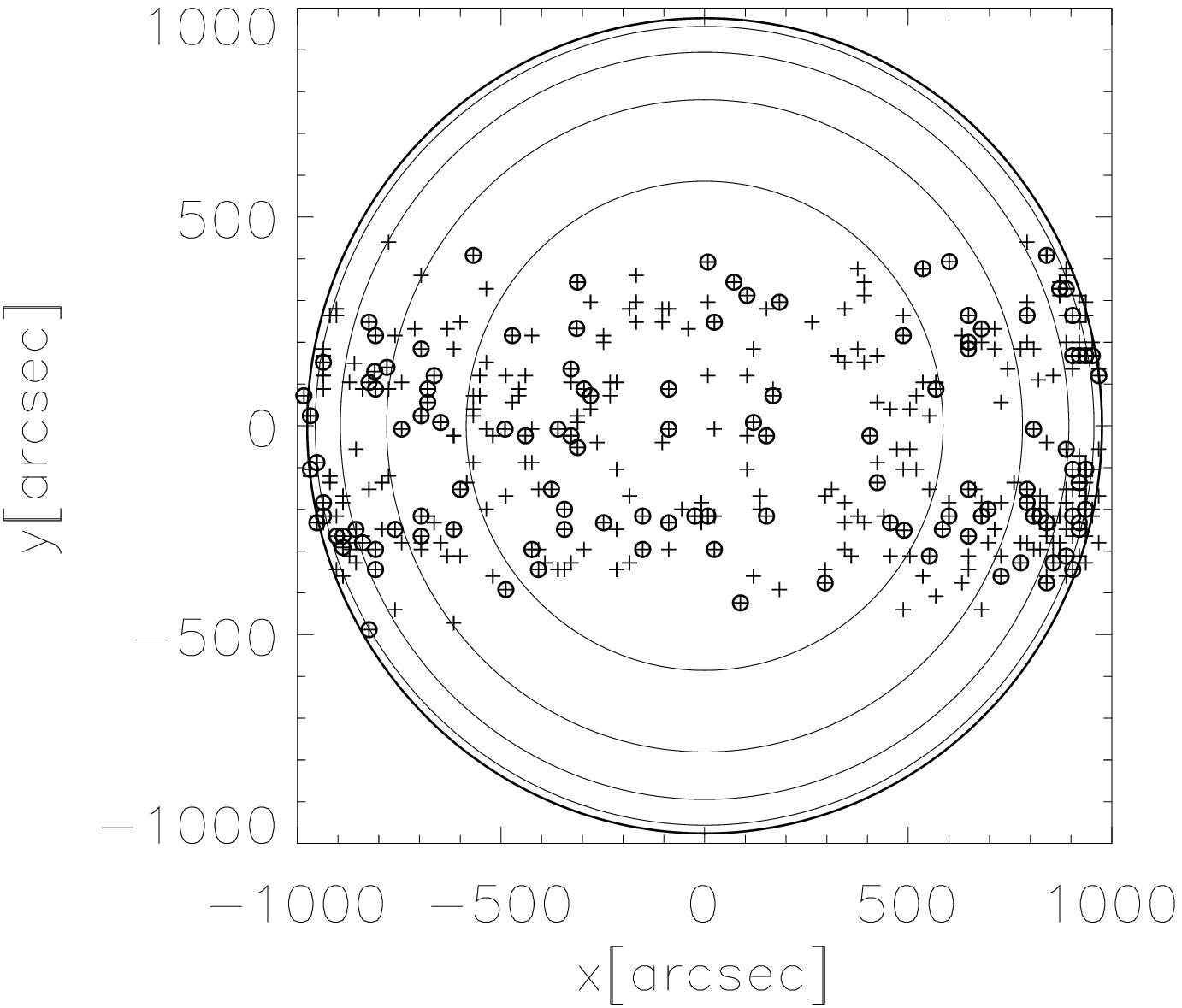}\\
\includegraphics[width=75mm]{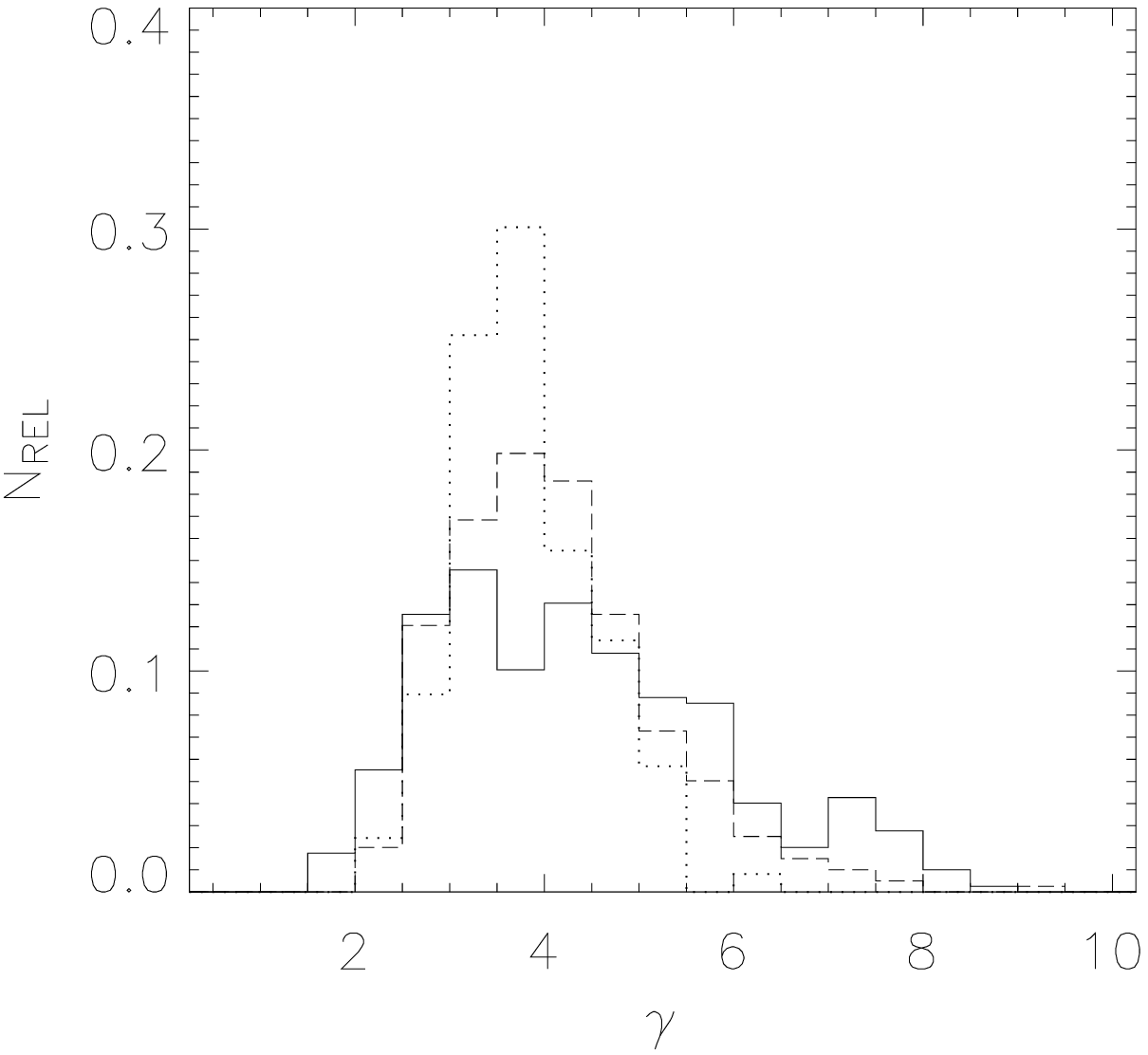}
\includegraphics[width=75mm]{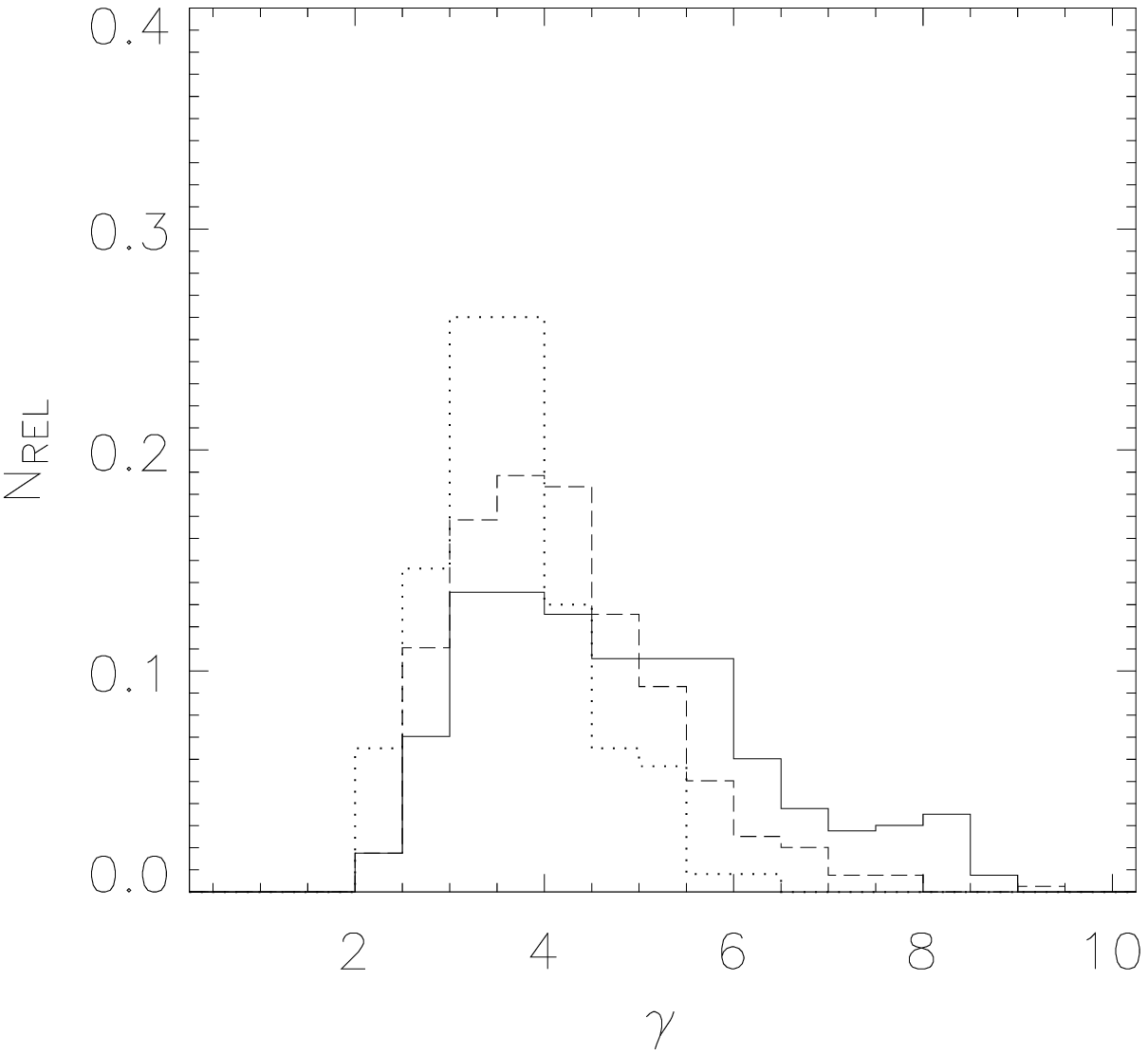}
\end{center}
\caption{Top: Positions of selected flares. 
Crosses indicate 398 
flares used for the centre-to-limb analysis of $\gamma_0$ and
$\gamma_1$, small circles correspond to 123 
flares for the $\gamma_2$ analysis (see text for details). 
Large circles indicate regions of $\mu=0.2$, 0.4, 0.6, and 0.8.
Middle and bottom: Relative occurrence of 
$\gamma_0$ (solid), $\gamma_1$ (dashed), and $\gamma_2$ (dotted)
from observed spectra and those corrected for isotropic albedo, respectively.}
\label{flare_pos}
\end{figure}

Hence, the above conditions select only events with significant flux above the roughly
estimated background in energies higher than the energy range where the spectral indices are determined.
This reduced our set to 398 events for $\gamma_0$ and $\gamma_1$ analysis
and to 123 events for $\gamma_2$ analysis -- see Figure~\ref{flare_pos}.
Note that the events are not uniformly distributed over $\mu$. The main reason
is that active regions and flares are located in a rather small range of
latitudes. Thus fewer events per unit $\mu$ are observed close to the limb than at the disc
centre. This also puts a constraint on the width of $\mu$ bins for
anisotropy analysis --  see Section~\ref{direc_sect} and Table~\ref{directivity}.

We also apply an isotropic $(\alpha(\mu)=1)$ 
albedo correction using Equation~(\ref{flux0}) and the approach 
discussed  in \citet{ko06} to determine spectral indices of albedo corrected spectra.
The relative occurrence of spectral indices of observed spectra and of those corrected 
for isotropic albedo is shown in Figure~\ref{flare_pos}. Note
the excess of events with $\gamma_0 < 2$ which is removed
after the isotropic albedo correction. The excess of $\gamma_0 > 7$ corresponds
to events with very steep spectra.
\section{Centre-to-limb variation}
\begin{figure}[ht!]
\begin{center}
\includegraphics[width=72mm]{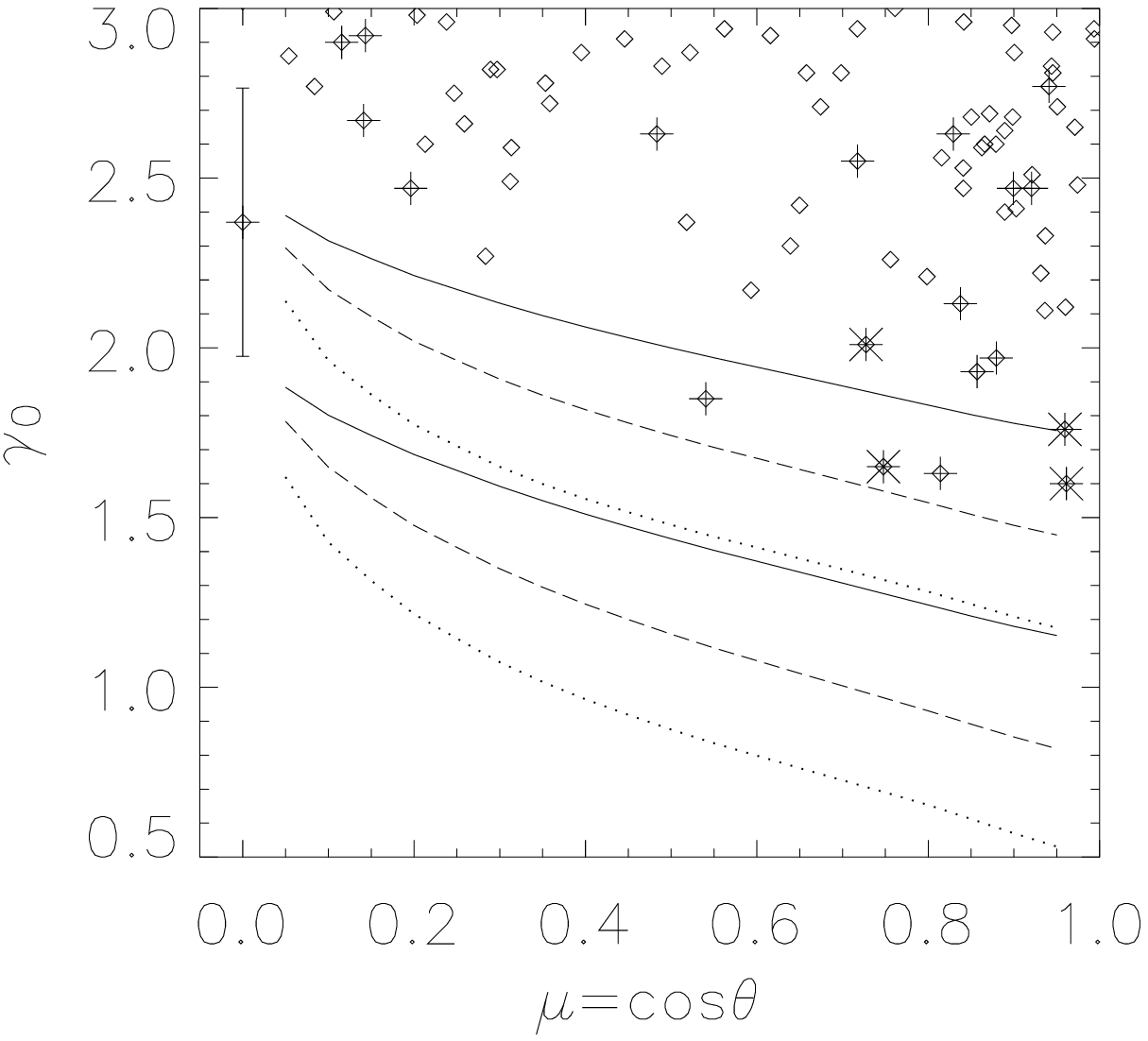}
\includegraphics[width=72mm]{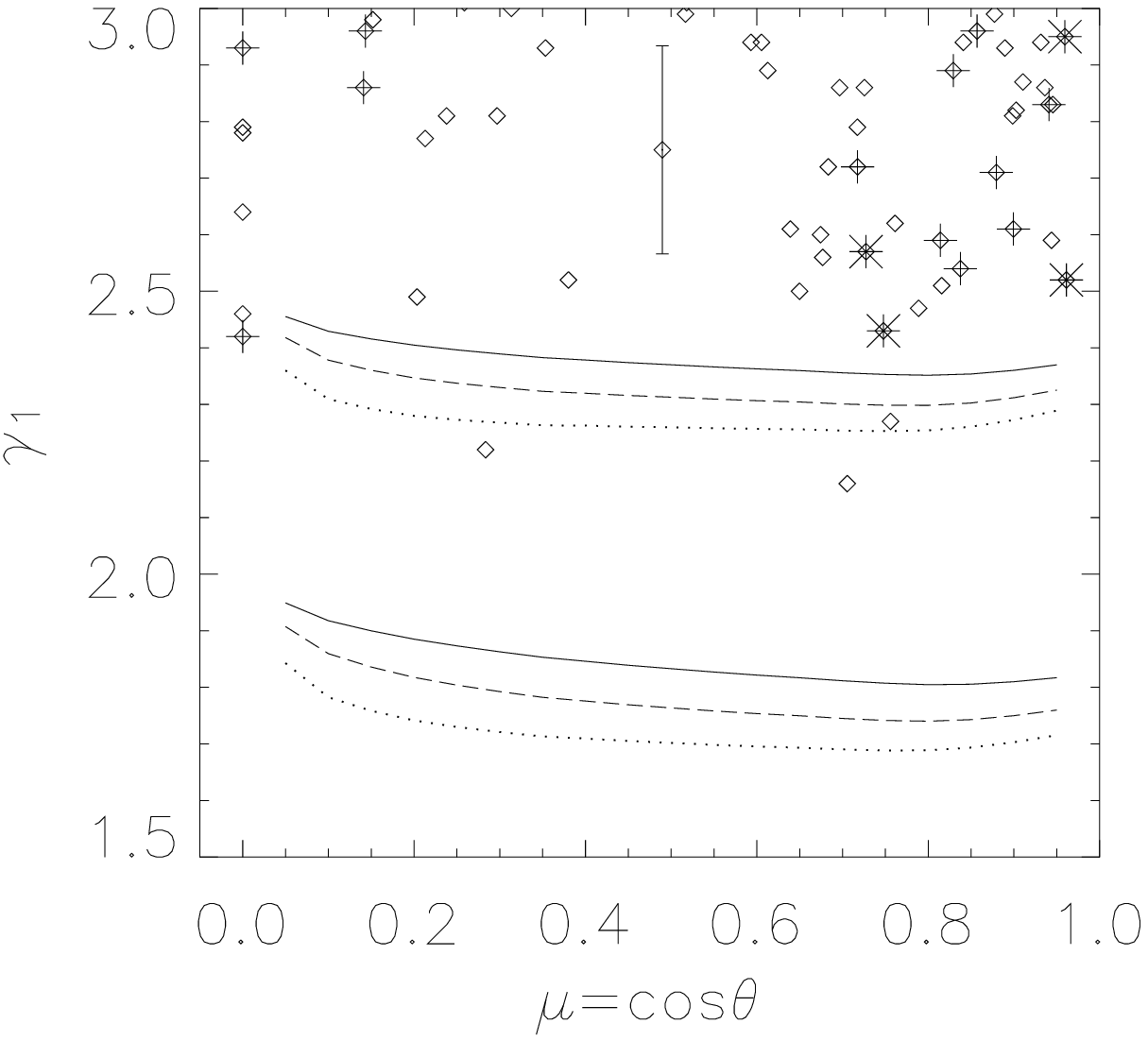}
\includegraphics[width=72mm]{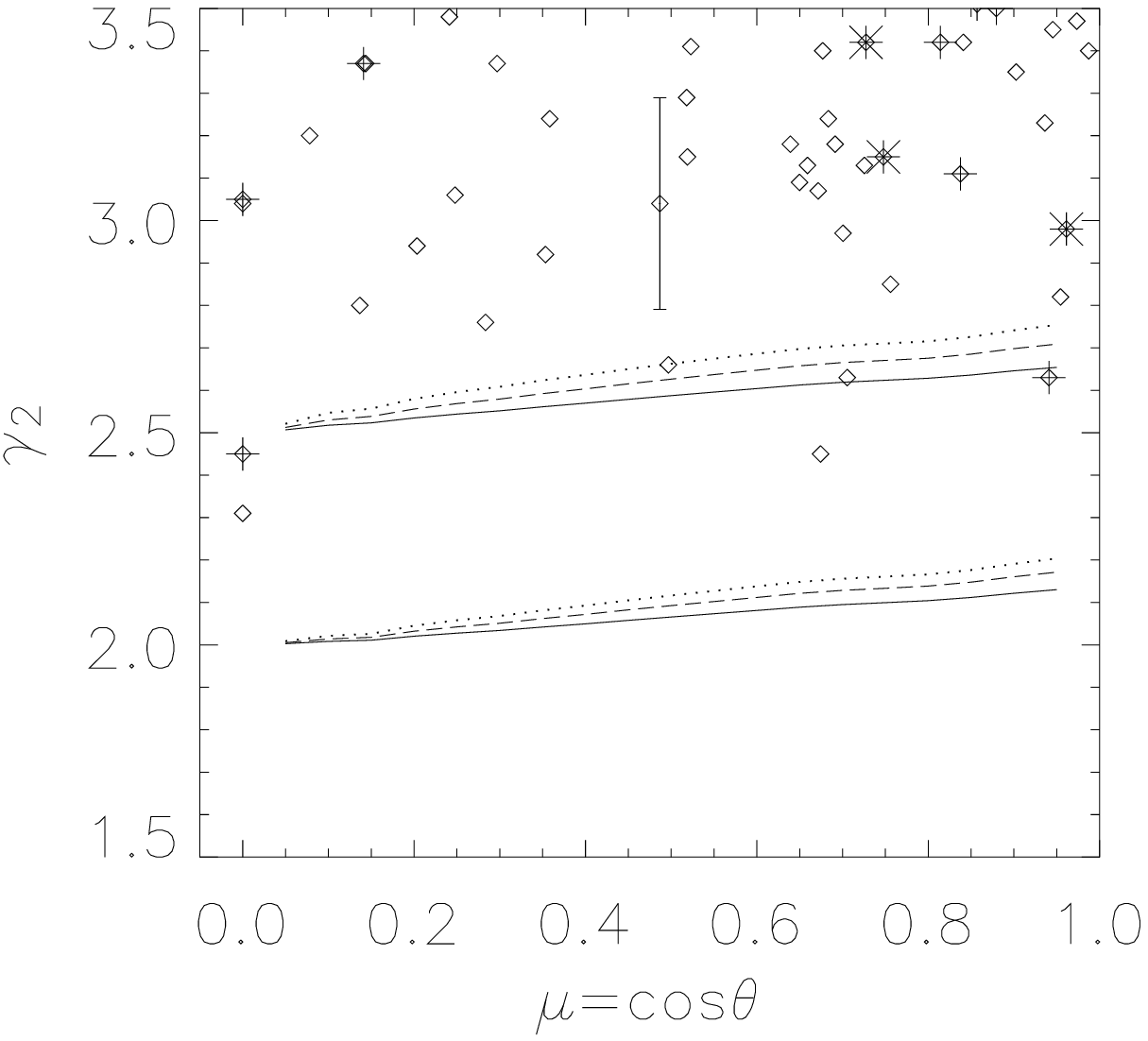}
\end{center}
\caption{Spectral indices $\gamma_0$, $\gamma_1$, and $\gamma_2$
versus cosine $\mu$ of heliocentric angle $\theta$. Vertical
error bars indicate average uncertainties on the values as
determined from single power-law fits. Lines show the predicted
dependency for single power-law primary spectra with $\gamma_\mathrm{p}$ =
2.0,2.5 and for $\alpha(\mu)=1,2,4$ (solid, dashed, and
dotted lines respectively), see also Section~\ref{direc_sect}. 
{\bf 
High values of spectral indices are not shown to emphasize their lower bounds.
Crosses represent flares for which $\bar{F}(E)$ was determined from
the total spectra. Flares with a dip in $\bar{F}(E)$ are denoted as stars
(see Section~\ref{cutoff}).} \label{statistics}}
\end{figure}

The resulting spectral indices $\gamma_0$, $\gamma_1$, and
$\gamma_2$ of the flares are shown Figure \ref{statistics}.
{\bf
There is no clear upper bound on $\gamma$ values (primary and observed spectral index can be 
arbitrarily large) and the albedo contribution to the observed spectrum
becomes negligible for steep primary spectrum - see Section~\ref{sect_albedo}. 
On the contrary, the primary spectral index 
for bremsstrahlung emission cannot be lower than 1.
Therefore,
}
for display purposes only values up to 3 (3.5 for $\gamma_2$) are shown;
events with larger spectral indices fill the whole range of $\mu$.

The spectral index $\gamma_0$ shows significant centre-to-limb
variation (see Figure~\ref{statistics}). Lower values of
$\gamma_0$ tend to be located closer to disc centre, forming an edge in
the $\gamma_0(\mu)$ distribution. There are no flares with $\gamma _0
<2$ located at $\mu <0.5$ ($\theta > 60^o$),  all such
being at $\mu\ge 0.5$. We do not see similar pattern at higher energies
(spectral indices $\gamma _1$ and $\gamma _2$) and the
Figure~\ref{statistics} shows no correlation with the position of
flares in these energy ranges.

As one can see from our albedo model with $\alpha\ne\alpha(\epsilon)$
(Figure~\ref{model}), the
strongest centre-to-limb variation due to  photospheric albedo is
expected at lower energies, i. e. for $\gamma_0$, and almost no
dependency is expected on  flare position at higher energies. The corresponding
theoretical model for single power-law primary spectra is shown by various
lines in Figure~\ref{statistics}. 

Figure~\ref{statistics} also shows that anisotropy
plays a significant role in centre-to-limb spectral variation only at $\gamma_0$
energies. Although we have no information about the primary spectra
and distribution of primary spectral indices of analysed events, 
it is clear from the figure that the albedo model
predictions, assuming single power-law primary spectra, are consistent
with the general variations of the observed spectral indices with $\mu$.
In order to put further constraint on the model, particularly at energies above 20 keV, 
spectral indices need to be determined with uncertainties smaller
than the predicted centre-to-limb variation -- see Figure~\ref{statistics}.

Following previous results \citep{ve87,da74,da77},
and to assess the change of spectral indices with $\mu$, we divide the sample
into two subsets; $\mu <0.5$, and $\mu \geq 0.5$ and for each
interval we calculate the mean values of $\gamma_0$, $\gamma_1$, and $\gamma_2$
of observed and isotropic albedo corrected photon spectra.
We then apply the Kolmogorov-Smirnov (K-S) test \citep{pr92}
to calculate the probability that
the distributions of spectral indices at $\mu < 0.5$ and $\mu \ge
0.5$ are drawn from the same parent distribution. 
The mean values
and the probabilities (see Table~\ref{meanvalues}) show the trend
expected for isotropic albedo. The mean $\bar {\gamma_0}$ value of ${\gamma_0}$
increases after the albedo correction is applied as one would
expect. The mean spectral indices $\bar {\gamma _1}$,
$\bar {\gamma _2}$ with and without albedo correction do not differ  
significantly -- i.e. are within  the uncertainties of the mean values.
The K-S probabilities show a similar trend,  the
distributions with the albedo component removed show larger
probabilities of being drawn from the same dataset. However, the
probability values for observed indices are too large to reject the
null hypothesis, partly because the albedo
contribution is smeared over too broad a range of $\mu$, $\Delta \mu=0.5$.  

Better results can be achieved by taking two more extreme distinct sets of
flares; one close to the limb with $\mu <0.1\ (\theta> 84\degr)$, where the albedo is
negligible, and disc centre events with $\mu \ge 0.9\ (\theta\le 25\degr)$ where the
albedo should be strongest. The value of K-S probability for
the $\gamma_0$ distributions drops to $0.9\%${\bf, allowing us to reject
the null hypothesis at 0.05 and 0.01 significant levels and conclude
that the $\gamma_0$ distributions are `significantly different'. 
} 

\begin{figure*}
\begin{center}
\includegraphics[width=65mm]{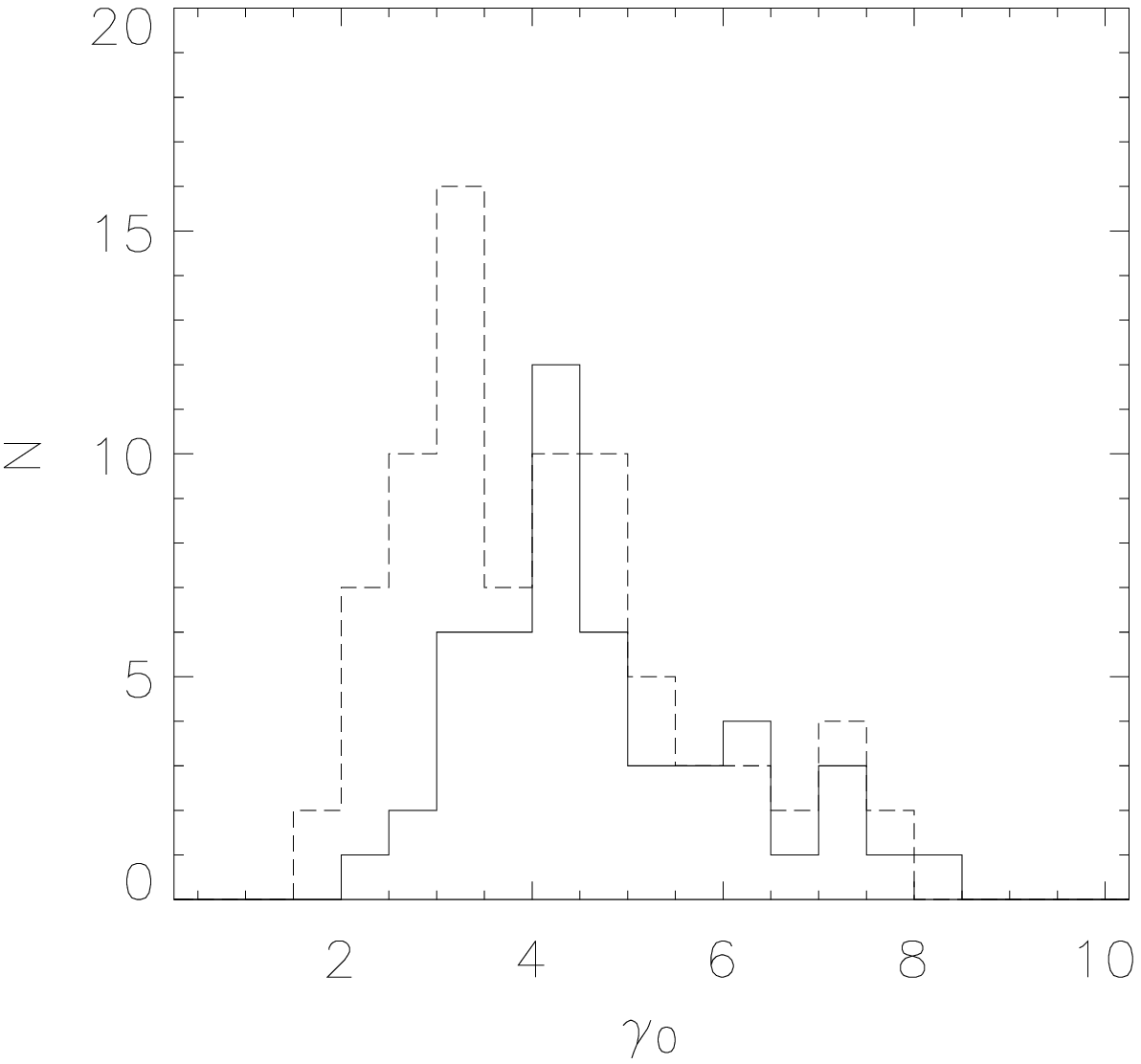}
\includegraphics[width=65mm]{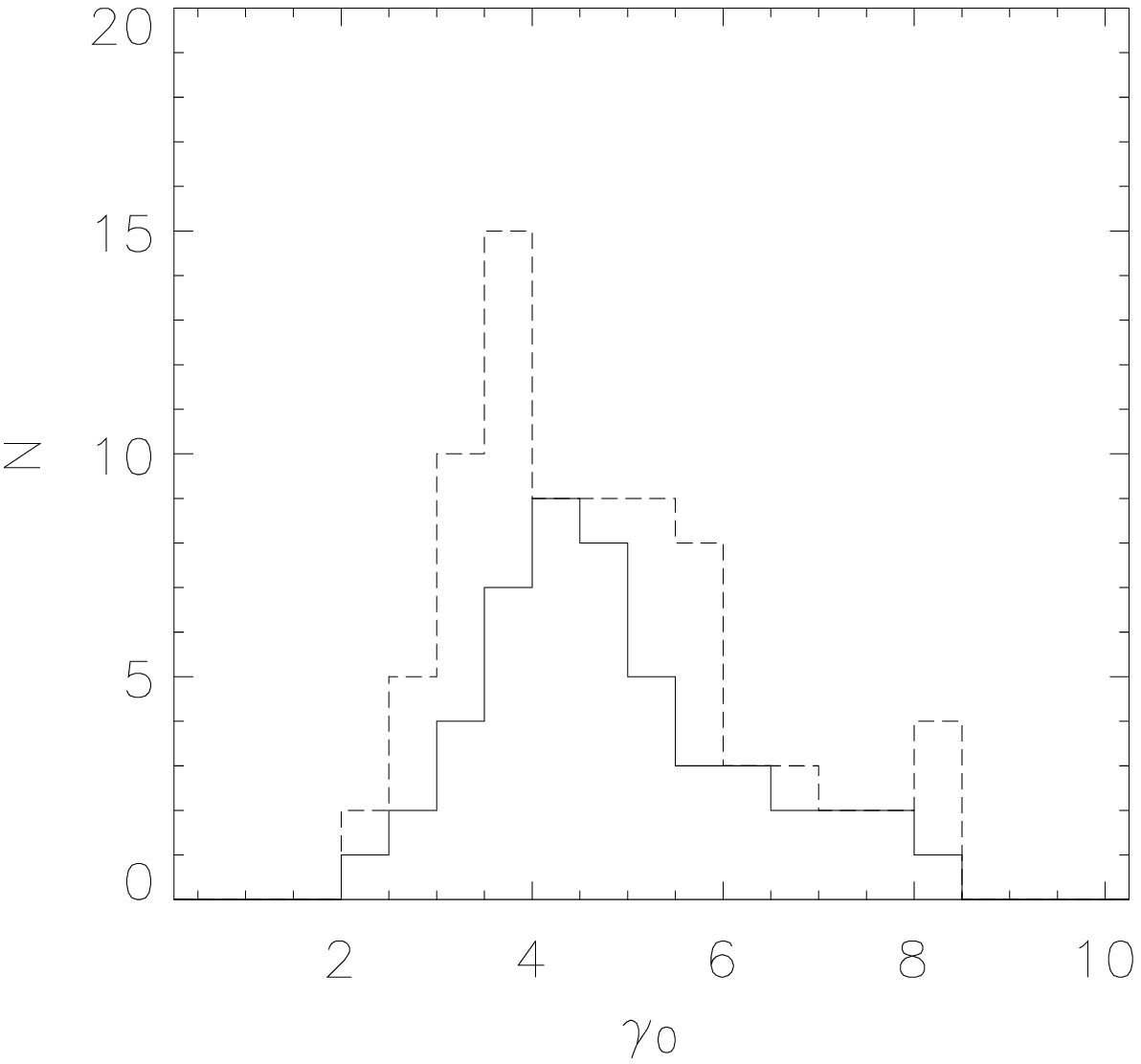}
\end{center}

\caption{Distributions of $\gamma_0$ for `centre' ($\mu > 0.9$, dashed line) and `limb'($\mu < 0.1$,
solid line) flares. The left plot corresponds to observed
$\gamma_0$, the right plot is for spectra corrected for isotropic
albedo, i.e. with directivity $\alpha = 1$.} \label{histograms}
\end{figure*}

After applying the isotropic albedo correction, the $\gamma_0$ distributions 
can no longer be considered as being different -- see Table~\ref{meanvalues} and 
the $\gamma_0$ distributions in Figure~\ref{histograms}.
However, the same conclusion cannot be reached for $\gamma_1$ and
$\gamma_2$ -- see Table~\ref{meanvalues}. It should be pointed out
that the number of events for $\gamma_2$ analysis for those $\mu$
bins is rather low -- see Table~\ref{meanvalues} -- so a larger
set of events will be needed to reach any conclusion. 

\begin{table*}
\begin{center}

\begin{tabular}{lccccc}
\hline
albedo corr.& $\mu$&$\bar{\gamma}_0$ (N)&$\bar{\gamma}_1$ (N)&$\bar{\gamma}_2$ (N)\\
\hline\hline
none&$\mu < 0.5$& $4.6\pm0.1\ (157)$&$4.2\pm0.1\ (157)$&$3.61\pm0.09\ (45)$\\
none&$\mu \ge 0.5$&  $4.3\pm0.1\ (241)$&$4.07\pm0.06\ (241)$&$3.88\pm0.09\ (78)$\\
\hline
K-S prob.&&0.2&0.2&0.2\\
\hline
$\alpha(\mu)=1$&$\mu < 0.5$& $4.8\pm0.1\ (157)$&$4.3\pm0.1\ (157)$&$3.5\pm0.1\ (45)$\\
$\alpha(\mu)=1$&$\mu\ge 0.5$& $4.8\pm0.1\ (241)$&$4.12\pm0.06\ (241)$&$3.6\pm0.1\ (78)$\\
\hline
K-S prob.&& 1.0 &0.2&0.5\\
\hline
none& $\mu < 0.1$ & $4.7\pm0.2\ (49)$&$4.4\pm0.2\ (49)$& $3.5\pm0.2\ (13)$\\
none& $\mu > 0.9$ & $4.2\pm0.2\ (81)$&$4.1\pm0.1\ (81)$& $3.9\pm0.1\ (29)$\\
\hline
K-S prob.&& 0.009  & 0.4 & 0.4\\
\hline
$\alpha(\mu)=1$& $\mu < 0.1$ & $4.8\pm0.2\ (49)$& $4.4\pm0.2\ (49)$& $3.4\pm0.2\ (13)$\\
$\alpha(\mu)=1$& $\mu > 0.9$ & $4.8\pm0.2\ (81)$  & $4.2\pm0.1\ (81)$& $3.6\pm0.1\ (29)$\\
\hline
K-S prob.&& 0.4 &  0.4 & 0.8\\
\hline
\end{tabular}
\end{center}

\caption{Mean spectral indices $\bar{\gamma _0}$, $\bar{\gamma
_1}$ and $\bar{\gamma _2}$ for all selected events at selected
ranges of $\mu$. K-S prob. corresponds to probability that a pair
of distributions comes from the same parent distribution. The
number of events in each $\mu$ bin is indicated in brackets.}
\label{meanvalues}

\end{table*}

\subsection{Directivity of X-ray emission}
\label{direc_sect}

The intensity of photons backscattered from the photosphere is
determined by the downward directed primary photon spectrum.
The larger the downward directivity of the source, the stronger is the albedo
contribution to the observed spectrum (see the predictions for 
$\alpha=1,2,4$ and
the model of a single power-law primary spectrum in
Figure~\ref{statistics}). We describe the downward versus toward observer flux ratio by
the anisotropy parameter $\alpha(\mu)$, see Equation~\ref{flux0}
(which we take to be independent of energy).

In order to assess the range of primary X-ray directivity consistent with the
data, we have assumed limb spectra, $I_\mathrm{o} (\epsilon,\mu < 0.1)$, 
to be true primary spectra toward the observer, 
$I_\mathrm{p} (\epsilon,\mu)\equiv I_\mathrm{o} (\epsilon,\mu < 0.1)$, 
and added to these the albedo component for various $\alpha(\mu)$. 
Using Equation~(\ref{flux0}), the modelled spectra $I_\mathrm{M}$  
in the observer direction $\mu$ take the form 
\begin{equation}
\label{flux0_add}
I_\mathrm{M}(\epsilon,\mu) = \left[{\bf 1} + 
\alpha(\mu){\bf G}(\epsilon,\epsilon',\mu)\right] I_\mathrm{o} (\epsilon',\mu < 0.1)\ .
\end{equation}

This approach does not decrease the signal-to-noise ratio 
of photon flux as in the case of albedo subtraction 
which reduces the photon flux but keeps the corresponding uncertainty at the same level
(uncertainty on the albedo contribution is set to zero.).

\begin{figure}
\begin{center}
\includegraphics[width=80mm]{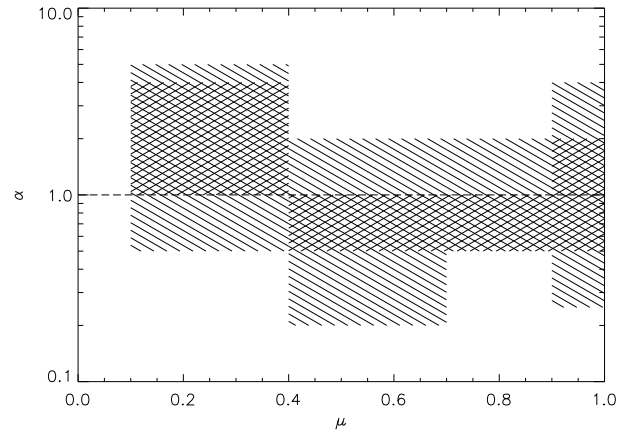}
\end{center}
\caption{Directivity at different $\mu$ determined from the distributions of $\gamma_0$ at the limb
and a given range of $\mu$.
Cross-hatched areas show the directivity for which the hypothesis that 
the observed distribution at a given range of $\mu$,
and the distribution $I_\mathrm{M}$ at the limb with added albedo contribution, 
are drawn from the same parent distribution
cannot be rejected at the 0.05 significance level; crossed areas correspond to the 0.01 significance level.
See also Table~\ref{directivity}. The dashed line shows the isotropic case, i.e. $\alpha(\mu)=1$.
}
\label{animu}
\end{figure}

The K-S test was then applied to distributions of spectral indices of the modelled $I_\mathrm{M}$ 
and observed $I_\mathrm{o}$ spectra
in several ranges of $\mu$. The width of the $\mu$ bins was chosen to contain 
approximately the same number of events.
Due to the non-uniform distribution of events with $\mu$, see Figure~\ref{flare_pos}, the width of the 
$\mu$ bins increases towards the limb (smaller $\mu$).
We regard the directivity $\alpha(\mu)$ as
consistent with the data if the K-S probability is larger 
than 0.05 or 0.01 significance level, i.e. when the null hypothesis that
two distributions of observed and modelled spectral index for given $\alpha(\mu)$ are drawn
from the same distribution cannot be rejected.
In the $\gamma_0$ energy range we find the ranges of
allowable directivity to be $0.5 < \alpha(\mu) < 4$ and
$0.2 < \alpha(\mu) < 5$, respectively -- see Table~\ref{directivity} and Figure \ref{animu}.
The results are also consistent the hypothesis that $\alpha(\mu)$ does not vary significantly with $\mu$
for the 0.05 and 0.01 significance levels.

In the energy range of  $\gamma_1$ no conclusions about anisotropy can be drawn
because high values of the K-S probabilities for the distribution of
$\gamma_1$ do not permit us to distinguish them, though, the results are not inconsistent
with the albedo model which does not predict significant variation
of $\gamma_1$ with $\mu$ and $\alpha$ in this energy range -- see Figure~\ref{statistics}.

Finally, the small number of suitable events at high energy prevented
us from analysing $\gamma_2$. On the basis of the
albedo model, see Figure~\ref{statistics}, we do not expect to
find a significant influence of the anisotropy in the $\gamma_2$ energy
range, but this prediction is not tested by the presented data.

\begin{table*}
\begin{center}

\begin{tabular}{cccccccccc}
&&&\multicolumn{7}{c}{K-S prob. for $\alpha(\mu)$}\\
\cline{3-10}
$\mu$&N&10&5&4&2&1&0.5&0.25&0.2\\
\hline\hline
$0.1 - 0.4$ &80 & $3\times 10^{-3}$&0.03 & 0.07 & 0.2 & 0.07 & 0.02 & $8\times 10^{-3}$ & $6\times 10^{-3}$\\
$0.4 - 0.7$ &88 & $9\times 10^{-6}$& $4\times 10^{-4}$ & $1\times 10^{-3}$ & 0.02 & 0.4 & 0.1 & 0.03 & 0.02\\
$0.7 - 0.9$ &100 &  $7\times 10^{-6}$&$2\times 10^{-4}$ & $4\times 10^{-4}$ & 0.02 & 0.3 & 0.04 & $7\times 10^{-3}$ & $7\times 10^{-3}$\\
$0.9 - 1.0$ &81 & $2\times 10^{-3}$& $9\times 10^{-3}$& 0.04 & 0.2 & 0.5 & 0.06 & $0.01$ & $4\times 10^{-3}$\\
\hline
\end{tabular}
\end{center}
\caption{K-S prob. for several values of directivities at different ranges of $\mu$ 
(N is the number of events in that range). The values correspond to
probability that the observed distribution of $\gamma_0$ at a given range of $\mu$ and the modelled distribution 
$I_\mathrm{M}$ of $\gamma_0$ determined from the events at the limb 
($\mu < 0.1$) with added albedo contribution for the given $\mu$ and directivity
$\alpha$ are drawn from the same parent distribution. 
}
\label{directivity}
\end{table*}

\section{Low-energy cutoff in $\bar{F}(E)$ demanded by the data}
\label{cutoff}

The survey presented in the this paper is also suitable for analysis
of flares where the hard X-ray spectrum may indicate suspicious features in $\bar{F}(E)$.

\begin{figure*}
\begin{center}
\includegraphics[width=60mm]{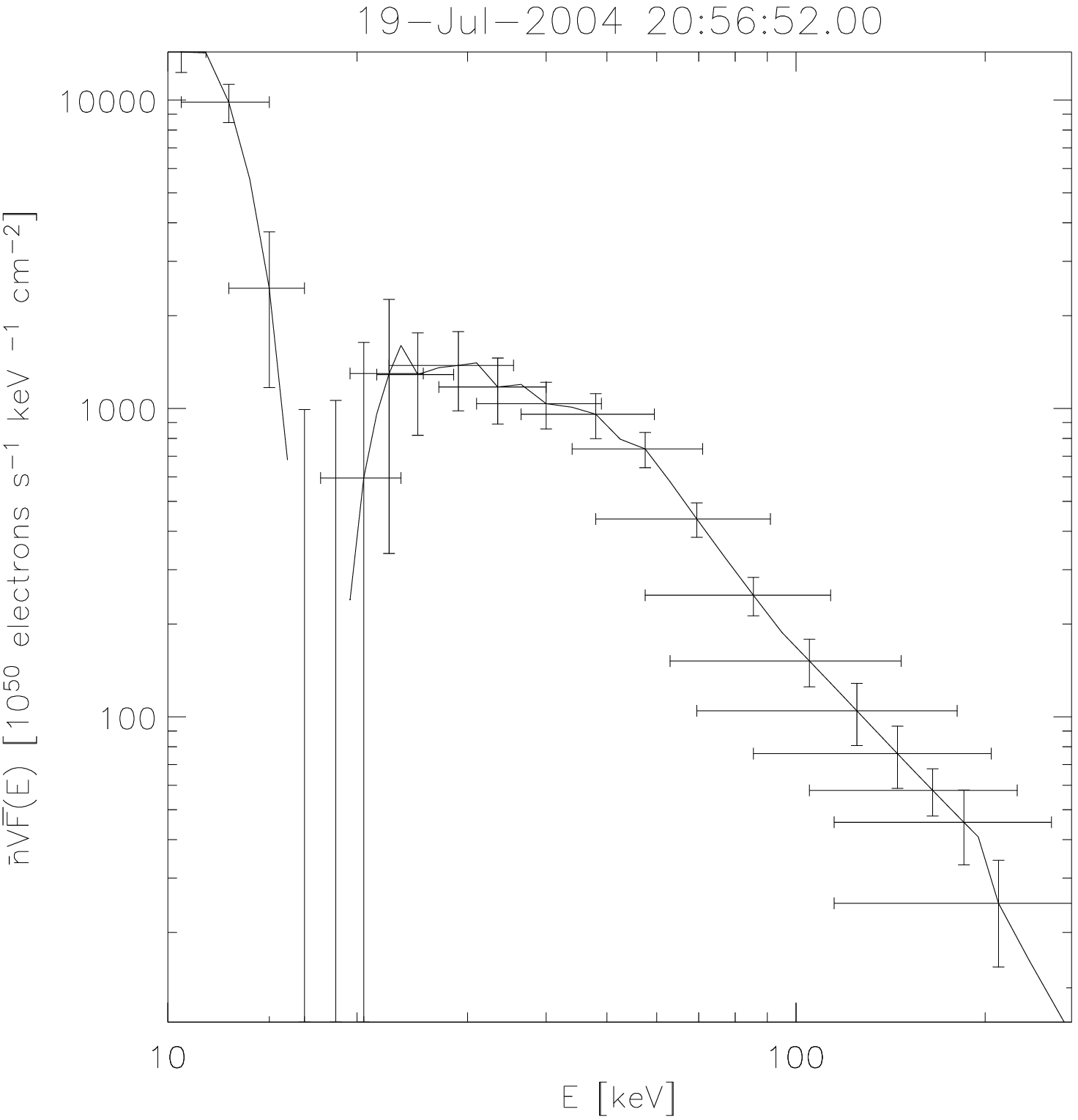}
\includegraphics[width=60mm]{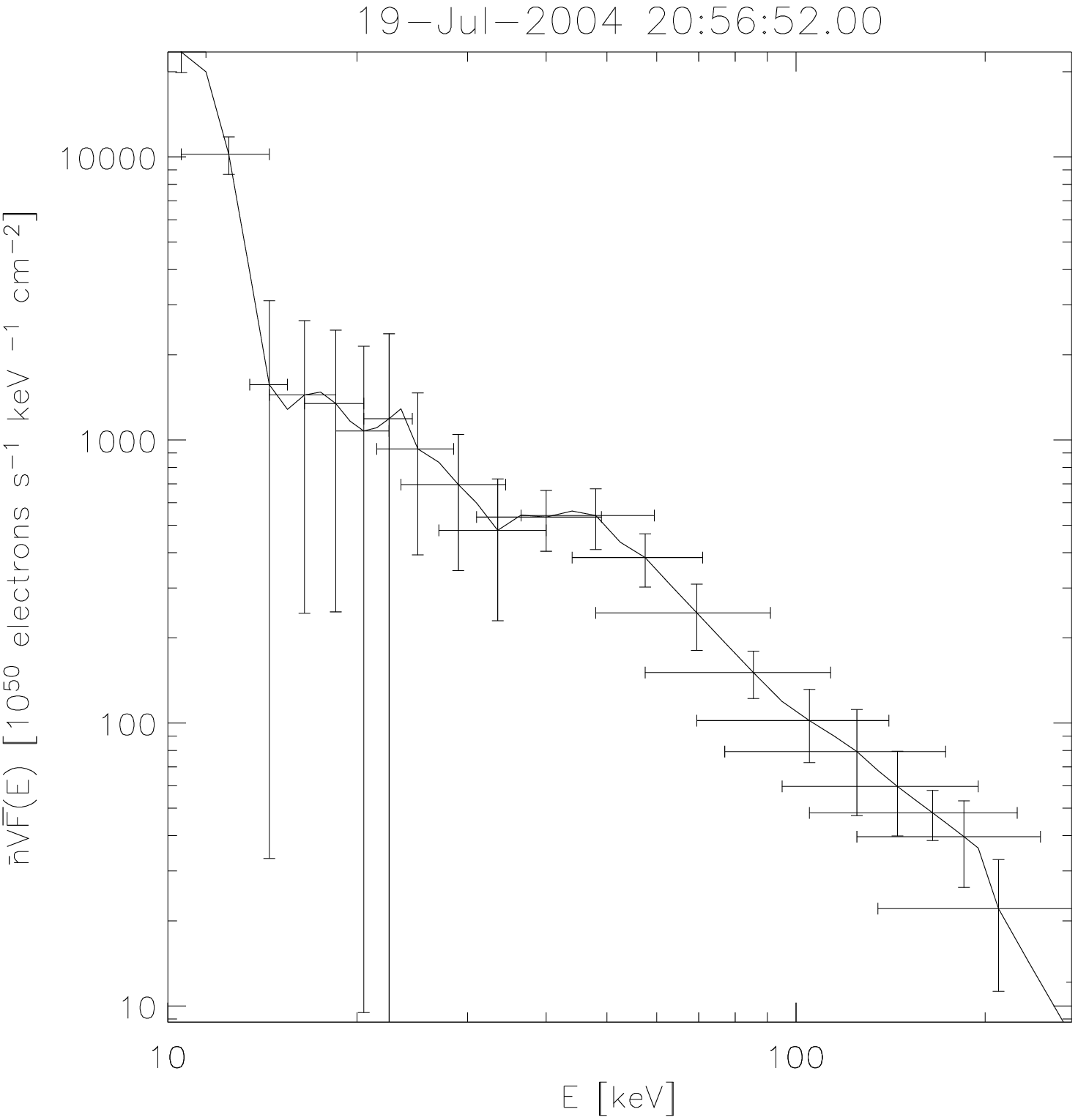}
\end{center}
\caption{$\bar{F}(E)$ corresponding to the total (left) and
primary (right) flare photon spectrum for an event close to the disc centre (19 Jul 2004).
Vertical error bars correspond to 3$\sigma$ uncertainties,
horizontal lines show the FWHM of energy resolution determined from
regularised inversion.} \label{fbar}
\end{figure*}

Using a regularised inversion technique \citep{ko04,sc06}, we
determined the mean electron flux distribution $\bar{F}(E)$ from
count spectra of several events which had either flat photon
spectra or a large change in local spectral index with energy. Such spectral
behaviour is expected for photon spectra either affected by the albedo -- see Figure~\ref{model} --
or produced by $\bar{F}(E)$ with an absence of electrons in same energy range.
We found several new cases which exhibit a statistically
significant dip in $\bar{F}(E)$ -- e.g. Figure~\ref{fbar}~(left).
The positions and spectral indices of events with such a dip are indicated in
Figure~\ref{statistics}. However, when the isotropic albedo
correction was applied (i.e. $\alpha (\mu)=1$), all such dips in $\bar{F}(E)$ were
removed - e.g. Figure~\ref{fbar}~(right). Our set of events includes the Sep 17, 2002 \citep{ko06} and Apr 25,
2002 flares \citep{ko06b} for which a statistically significant
dip in $\bar{F}(E)$ at energies below 30 keV has been reported,
neglecting albedo. 

The albedo contribution in events close to the limb should be
negligible. Therefore, features in $\bar{F}(E)$ for events there could not be
ascribed to the distortion of the photon spectra by albedo.
$\bar{F}(E)$ was determined for a number of events close to the limb
which showed flattening of the photon spectra. 
None of these events close to the solar limb ($\mu <0.5$)
was found to have a statistically significant dip in $\bar{F}(E)$,
but revealed only some flattening of $\bar{F}(E)$. 
{\bf All events with determined $\bar{F}(E)$ are indicated in Figure~\ref{statistics}.}

\section{Discussion and conclusions}
The high energy resolution of RHESSI down to $\sim 1$~keV and a large
database have allowed us to study statistically  398 
events not affected by pulse pile-up but having sufficient count rate to
measure local spectral indices in three different energy bands.
Combining these with flare position measurements we analysed
centre-to-limb variation of hard X-ray flare spectra.
The data clearly show a distinct variation of low energy spectral indices 
with heliocentric angle. 
This dependency is strong for spectral index in the
range 15~-~20 keV but very weak for higher energies. 
The mean values of low energy index $\gamma_0$ are decreasing for larger $\mu$ (smaller heliocentric
angles). We have compared centre-to-limb spectral variations with the predictions of  albedo
induced spectral index change \citep{ko06} for primary spectra assumed to be power-law
and found the behaviour of mean spectral indexes $\bar \gamma_0$, $ \bar \gamma_1$, and $\bar
\gamma_2$ (energy ranges 15~-~20~keV, 20~-~35~keV, and 35~-~50~keV)
to be consistent with the spectral change due to Compton backscattering albedo.
For $\gamma_1$ and $\gamma_2$ at higher energies there is very little
constraint information in the data. 

The intensity of backscattered photons is determined by intensity of downward beamed photons,
so we can estimate the ratio of downward/observer directed
photons. Although our data set contains large number of events
of photon spectra with high energy resolution, it can be only used to
test the hard X-ray anisotropy below 20 keV. Even for an anisotropic primary source, the albedo induced
spectral change and centre-to-limb variation are significant only below 20 keV -- see Figure~\ref{model}
and~\ref{statistics} (which applies for a primary power-law photon spectrum).
The inferred anisotropy ratio,  $0.2 < \alpha(\mu) < 5$, (Figure~\ref{animu}) in the 15~-~20~keV range,
and especially its large spread, allows no clear conclusion about the anisotropy.
Moreover, at these energies anisotropy of electron beam emission can be masked by
isotropic thermal emission, thus making the determination of anisotropy from spatially
integrated spectra difficult.

The  value of anisotropy ratio obtained is consistent with the predictions of 
hard X-ray emission  of downward collimated beam
and cannot reject even quite strongly beamed cases. 
For example, the detailed model of \citet[][ Fig.4]{le83} predicts
$\alpha\la 3$ at 22~keV  for disc centre events.
On the other hand, the determined anisotropy $\alpha(\mu)$ 
is also consistent with isotropic X-ray emission
and lends support to quite isotropic electron distribution when put together with
the \citet{ko06c} analysis of detailed spectra of individual events.
The anisotropy parameter $\alpha(\mu)$ 
was also found not to need significant dependence on $\mu$ -- see Figure~\ref{animu} --
indicating that the beam X-ray emission need not strongly vary in the upward hemisphere (different
values of $\mu$ probe different viewing angles of upward emission).
We stress that with $\alpha(\mu)$ 
we measure the anisotropy of downward directed photons rather than upward directed photons at different angles
as was done in most previous studies of solar flare directivities.

The data analysed in the paper, as well as previous
published results \citep{ni90,fa97,ka05b,ko06b} describe events with observed photon spectra 
which are flat at low
energies. Such behaviour suggests a low-energy cutoff in the mean electron spectrum as derived from such
observed photon spectra. We show that all such flares were located close to the solar disc centre ($\mu
> 0.5$ ($\theta < 60\degr$)), where the albedo component must be properly taken into account.
\citet{ni90} reported that two flares observed with SMM on 15 October 1981 at 4:43 UT 
and on 29 March 1980 at 9:18 UT 
had $\gamma \sim 2$ and $\gamma = 1.9$, respectively, 
both flares being located not far from disc centre ($\mu\ge 0.6$).
\citet{fa97} observed a few flares with very flat observed hard X-ray
photon spectra from the same active region near the
disc centre ($\mu=1.0$). 
We note that \citet{zh03,zh04} studied the combined effect of the low-energy
cutoff and albedo contribution to explain very flat photon spectra observed by Yohkoh/HXT
and concluded that both effect could cause flattening of photon spectra.
However, to account for albedo they used {\it integral} reflectivity (angle-averaged)
which is not applicable to centre-to-limb studies.

We have applied an isotropic albedo correction to the flares
in our survey and have found this makes the alleged  low-energy cutoff (or a `dip')
in electron spectra statistically insignificant. 
This survey strongly suggests
the hypothesis that the low-energy cutoff inferred in the mean electron flux
distribution is an artefact of the photospheric albedo.
Hence, we can confidently confirm our earlier conclusion
\citep{ko06} that low energy cutoff is a feature connected with
albedo and  {\it not} a physical property of the mean electron
distribution. This result can substantially change the estimated flux and total
number of electrons accelerated in a flare. Therefore, we strongly suggest that
the albedo component must be considered in determination of electron beam properties from spatially
integrated solar hard X-ray spectra.

\begin{acknowledgements}
{\bf We are grateful to H.~S. Hudson and C.~M. Johns-Krull for their valuable comments.}
JK acknowledges the financial support
of a Royal Society Incoming Short Visit grant and a PPARC RA post
to visit University of Glasgow in 2005 and 2006, and of grants 205/04/0358 and
205/06/P135 of the Grant Agency of the Czech Republic, and of
research plan AV0Z10030501 of the Astronomical Institute AS CR. EPK and JCB
acknowledge the financial support of a PPARC Advanced Fellowship
and Rolling Grant respectively.
This work was also supported by the Visiting Science Program at
International Space Science Institute, Bern, Switzerland.
\end{acknowledgements}

\bibliographystyle{aa}
\bibliography{pub,albedo_ref}
\end{document}